\documentclass{ws-ijmpaXXX}

\usepackage{pifont}  %--> non-standard?
\newcommand{\ipb}{\ensuremath{\mathrm{~pb^{-1}}}}

\newcommand{\GeV}{\ensuremath{\mathrm{~Ge\kern -0.1em V}}}
\newcommand{\GeVc}{\ensuremath{\mathrm{~Ge\kern -0.1em V\!/}c}}
\newcommand{\MeVc}{\ensuremath{\mathrm{~Me\kern -0.1em V\!/}c}}
\newcommand{\GeVcc}{\ensuremath{\mathrm{~Ge\kern -0.1em V\!/}c^2}}
\newcommand{\MeVcc}{\ensuremath{\mathrm{~Me\kern -0.1em V\!/}c^2}}
\newcommand{\MeV}{\ensuremath{\mathrm{~Me\kern -0.1em V}}}

\begin{document}    %CDF7235

\markboth{The $X(3872)$ at CDF~II --- G. Bauer~~~[DPF04]}
{The $X(3872)$ at CDF~II --- G. Bauer~~~[DPF04]}

%%%%%%%%%%%%%%%%%%%%% Publisher's Area please ignore %%%%%%%%%%%%%%%
%
\catchline{}{}{}{}{}
%
%%%%%%%%%%%%%%%%%%%%%%%%%%%%%%%%%%%%%%%%%%%%%%%%%%%%%%%%%%%%%%%%%%%%

\title{The $X(3872)$ at CDF~II}

\author{\footnotesize G. Bauer\\
(Representing the CDF~II Collaboration)
%\footnote{
%Typeset names in 8 pt Times Roman, uppercase. Use the footnote to 
%indicate the present or permanent address of the author.}
}

\address{Laboratory of Nuclear Science, Massachusetts Institute of Technology, \\
77 Massachusetts Avenue, Cambridge, MA 02139, USA
%\footnote{State completely without abbreviations, the 
%affiliation and mailing address, including country and e-mail address. 
%Typeset in 8 pt Times Italic.
}
%\\
%bauerg@fnal.gov}

\maketitle

%\pub{Received (Day Month Year)}{Revised (Day Month Year)}

\begin{abstract}
Last year's  $X(3872)$ discovery was confirmed with
the CDF II detector in $\bar{p}p$ collisions.
We measure its mass
to be $3871.3\pm0.7\pm0.4\MeVcc$.
The source of $X$-mesons in the large CDF sample is resolved
by studying their vertex displacement.
We find  $16.1\!\pm\!4.9\!\pm\!2.0\%$ of our $X$-sample
comes from decays of $b$-hadrons, and the remainder
from prompt sources: either direct production 
or by decay of (unknown) short-lived particles.
The mix of production sources is similar to that observed
for the $\psi(2S)$ charmonium state.
%
%The abstract should summarize the context, content and conclusions of
%the paper in less than 200 words. It should not contain any references
%or displayed equations. Typeset the abstract in 8 pt Times Roman with
%baselineskip of 10 pt, making an indentation of 1.5 pica on the left
%and right margins.

\end{abstract}

%\section{Introduction}

%\raggedright

~\\	
At last year's Lepton-Photon Symposium Belle
announced discovery of a charm\-on\-ium-like state,\cite{BelleX,BelleXPRL}  
$X(3872)$,
in $B^+ \!\rightarrow\! K^+ J/\psi\pi^+\pi^-$.
%exclusively reconstructed $B$-decays.
CDF %\cite{XCDFQuarkOnia}  
quickly confirmed $X \!\rightarrow\! J/\psi\pi^+\pi^-$.\cite{XCDF}
A natural interpretation of the $X$
%the $X(3872)$ 
is the $^3D_2$ of $c\bar{c}$,
but this is contrary to expectations.
The  $^3D_2$ is thought to be significantly lighter
($\sim\!3830\MeVcc$);
and Belle failed to detect decays to $\chi_{c1} \gamma$, 
which should be prominent for $^3D_2$.
More circumstantial is the expectation of a relatively flat 
dipion mass ($M_{\pi\pi}$) distribution for $D$-states,\cite{Yan}
whereas Belle found high masses preferred---possibly 
consistent with the (isospin violating) decay to $J/\psi\rho^0$.
These difficulties, coupled with the proximity of the $X(3872)$ 
to the $D^0\overline{D}{^{*0}}$-threshold,  prompted
speculation that the $X$ may be a  $D^0$-$\overline{D}{^{*0}}$
``molecule.''
Whether this is the case, or the $X$ is ``only'' a $c\bar{c}$-state 
in conflict with current theoretical models,
the $X$ is an interesting object of study.\cite{XLit}

\begin{figure}[t]
\centerline{\psfig{file=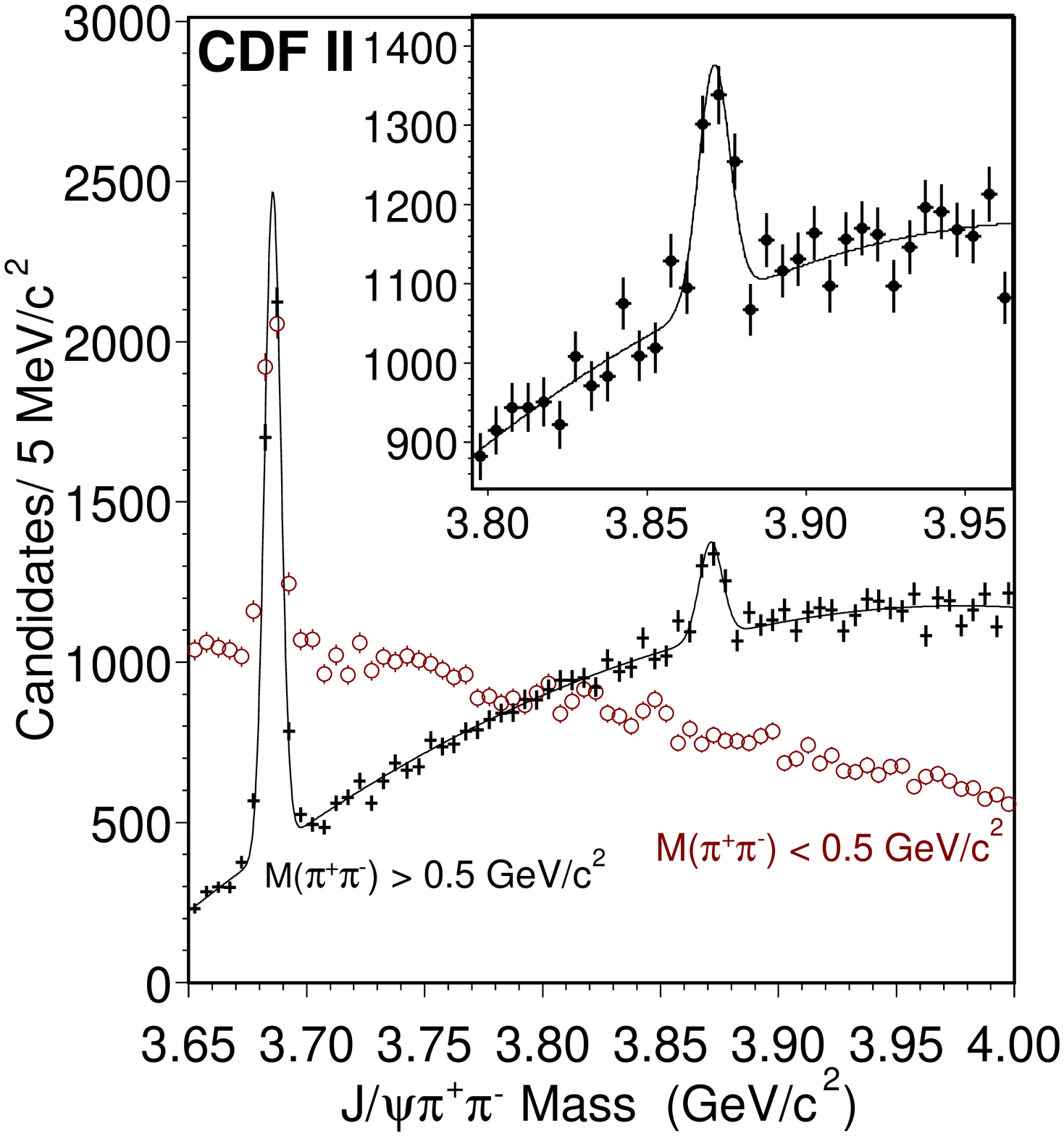,width=2.15in}
\begin{minipage}[b]{0.55\textwidth}
\hspace*{-0.15cm}\mbox{\psfig{file=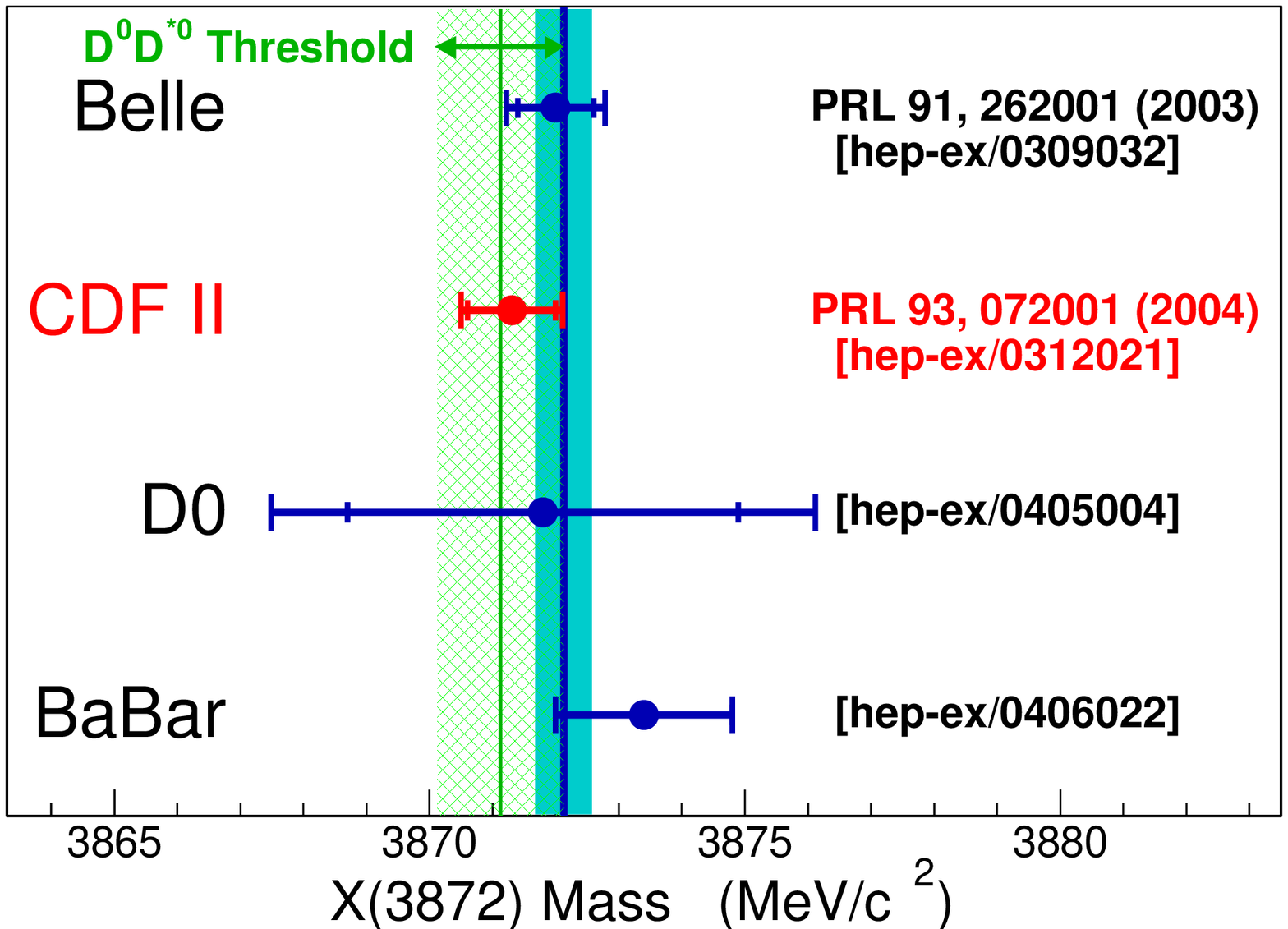,width=2.95in}}\\
~
\end{minipage}
           }
\vspace*{8pt}
\caption{
         {\bf LEFT:} The $J/\psi\pi^+\pi^-$ mass distribution for
           $M_{\pi\pi}\!<\!500$ and $>\!500\MeVcc$ subsamples.
         {\bf RIGHT:} Summary of $X$-mass measurements compared
                      to the  $D^0\overline{D}{^{*0}}$ threshold.
\label{Fig:XCDF}
}
\end{figure}

CDF~II\cite{CDFII} is a general purpose detector at Fermilab's $\bar{p}p$ collider.
We use 220\ipb\ of $\mu^+\mu^-$ triggers, 
yielding a clean 
$J/\psi$ sample.
%$J/\psi\!\rightarrow\!\mu^+\mu^-$ sample.
%The $X(3872)$ was quickly confirmed by CDF~II\cite{XCDFQuarkOnia,XCDF} 
%via the same $J/\psi\pi^+\pi^-$ mode.
Aside from technical cuts,
kinematic and spatial cuts
suppress large backgrounds 
from $J/\psi$'s plus random tracks.
The main cuts are:
a maximum number of $J/\psi\pi\pi$ candidates/event,
$p_T(J/\psi) \!>\!4\GeVc$,
$p_T(\pi)\!> \! 400\MeVc$, and
$\Delta R \!\equiv\!$ $ \sqrt{(\Delta\phi)^2+(\Delta\eta)^2}   \!<\! 0.7$
for each pion,
where $\Delta\phi$ ($\Delta\eta$) is the azimuthal
(pseudorapidity) difference of the pion with respect 
to the $J/\psi\pi\pi$.
With these cuts a significant $X$-signal is revealed.\cite{XCDF}
%These cuts are sufficient to reveal a significant $X$-signal.\cite{XCDF}
Here, however, we show in Fig.~\ref{Fig:XCDF} the results split up 
into $M_{\pi\pi}\!<\!500$ and  $>\!500\MeVcc$ subsamples.
No $X$-signal is apparent for low $M_{\pi\pi}$, 
supporting Belle's observation of high-mass decays.

Using the high-$M_{\pi\pi}$ sample, the $X$-mass is  
%$3871.3\!\pm\!0.7\!\pm\!0.4\MeVcc$.
$3871.3\pm0.7\pm0.4\MeVcc$.
Also shown in Fig.~\ref{Fig:XCDF} are masses from other experiments,
and the average compared to the  $D^0\overline{D}{^{*0}}$ threshold.
%This remarkable proximity fosters a  $D^0\overline{D}{^{*0}}$-molecular
%interpretation.
The near equality helps fuel  molecular-$D^0\overline{D}{^{*0}}$
speculations.

%psi sample with all combinatorics -> backgrount ~8x larger than actual selection

% \begin{figure}[t]
% \centerline{\psfig{file=avemass.eps,width=2.6in}%~~\psfig{file=PRL_fig2.eps,width=2.1in}
%            }
% \vspace*{8pt}
% \caption{{\bf LEFT:} $X$ mass summary
% \label{Fig:XMass}
% }
% \end{figure}
% 

From Belle's observation,
$B$-mesons are a significant source of $X$'s.
This raises some questions:
Is the CDF sample only from  $b$-hadrons?
%versus direct production?
If not, is direct $X$ production different from charmonium?
The technique of separating $b$-decay feeddown
from prompt sources is well established.\cite{Psi2SCDF}
Since $X$-decay is not weak, it is too rapid
to leave a displaced vertex.
If, however, it is produced by a (boosted) $b$-decay, 
the $X$  will be displaced due to the $b$-lifetime.
We measure the transverse $X$-displacement,
$L_{xy}$, and convert it to an ``uncorrected'' proper-time:
$ct = M\!\cdot\! L_{xy}/p_T$.
This is not the true proper-time of the $b$-decay because
$M$ and $p_T$ are only for the $J/\psi\pi^+\pi^-$.
% This distorts the ``lifetime'' scale, but not the 
%  {\it fraction} which is long-lived.

We use the same $X$-sample as above, but now impose additional
cuts related to the $Si$-vertex tracker,
mainly to demand  $\sigma(L_{xy})\!<\!125\,\mu$m
and have good beamline information.
The sample is reduced by $\sim\!15\%$.
An unbinned likelihood fit is performed
simultaneously over the $ct$ and mass of the candidates.
The signal is modeled by a Gaussian in mass; and 
for the $ct$-distribution,
a resolution smeared exponential for the long-lived component and
by the resolution function for the prompt.
The background model uses a quadratic polynomial for mass,
and resolution function for the prompt and {\it three}
resolution smeared exponentials---one for the negative-$ct$ tail
and two for the positive.
The resolution function consists of two Gaussians.
%  whose widths
% are the measured candidate's $\sigma(ct)$ multiplied 
% by two respective global scale factors, which in turn float in the fit.

\begin{figure}[t]
%\centerline
%\begin{minipage}[b]{0.60\textwidth}
%{\psfig{file=Blessed2S-LifeProj2.5.eps,width=2.6in}
%           }
%\end{minipage}
%\vspace*{8pt}
%\vspace*{-1.2cm}
\begin{minipage}[b]{0.42\textwidth}
\caption{Projection of $\psi(2S)$ likelihood fit onto the
         uncorrected proper-time distribution for the full PDF, and its breakdown
         into signal (shaded) and background (hatched) classes.
         Signal and background are further separated into
         prompt and long-lived components.
         The projections are
         for candidates within $\pm2.5\sigma$ of the $\psi(2S)$ mass
         in order to be reflective of its signal-to-background ratio. 
         The fit actually spans the mass range 3640-3740\MeVcc.
\vspace*{16pt}
\label{Fig:2SCDF}
}
\end{minipage}\hfill
\begin{minipage}[b]{0.57\textwidth}
{~~~~\psfig{file=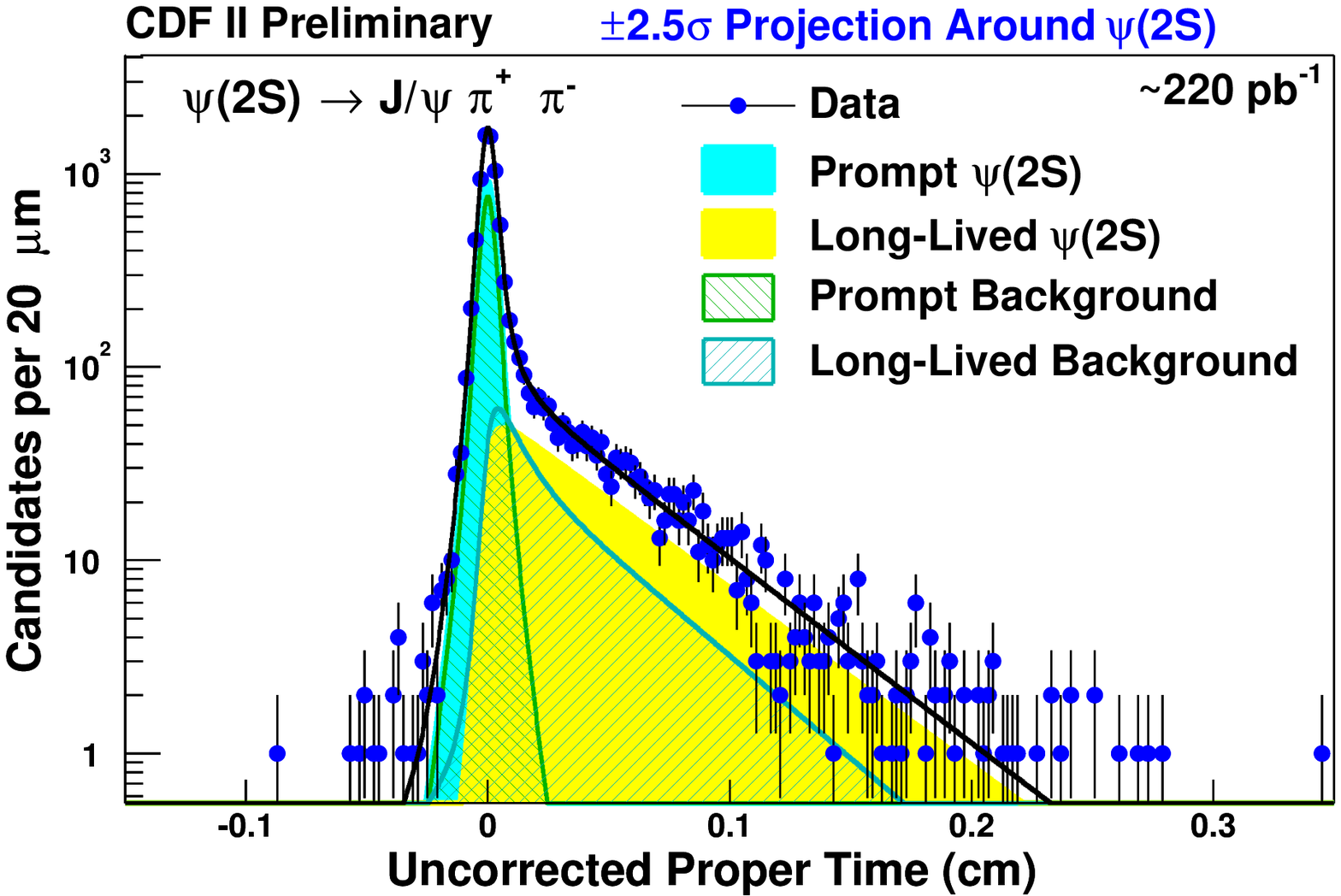,width=\textwidth}   %width=2.9in}
           }
\end{minipage}
\vspace*{-18pt}
\end{figure}

\begin{figure}[t]
\centerline{\psfig{file=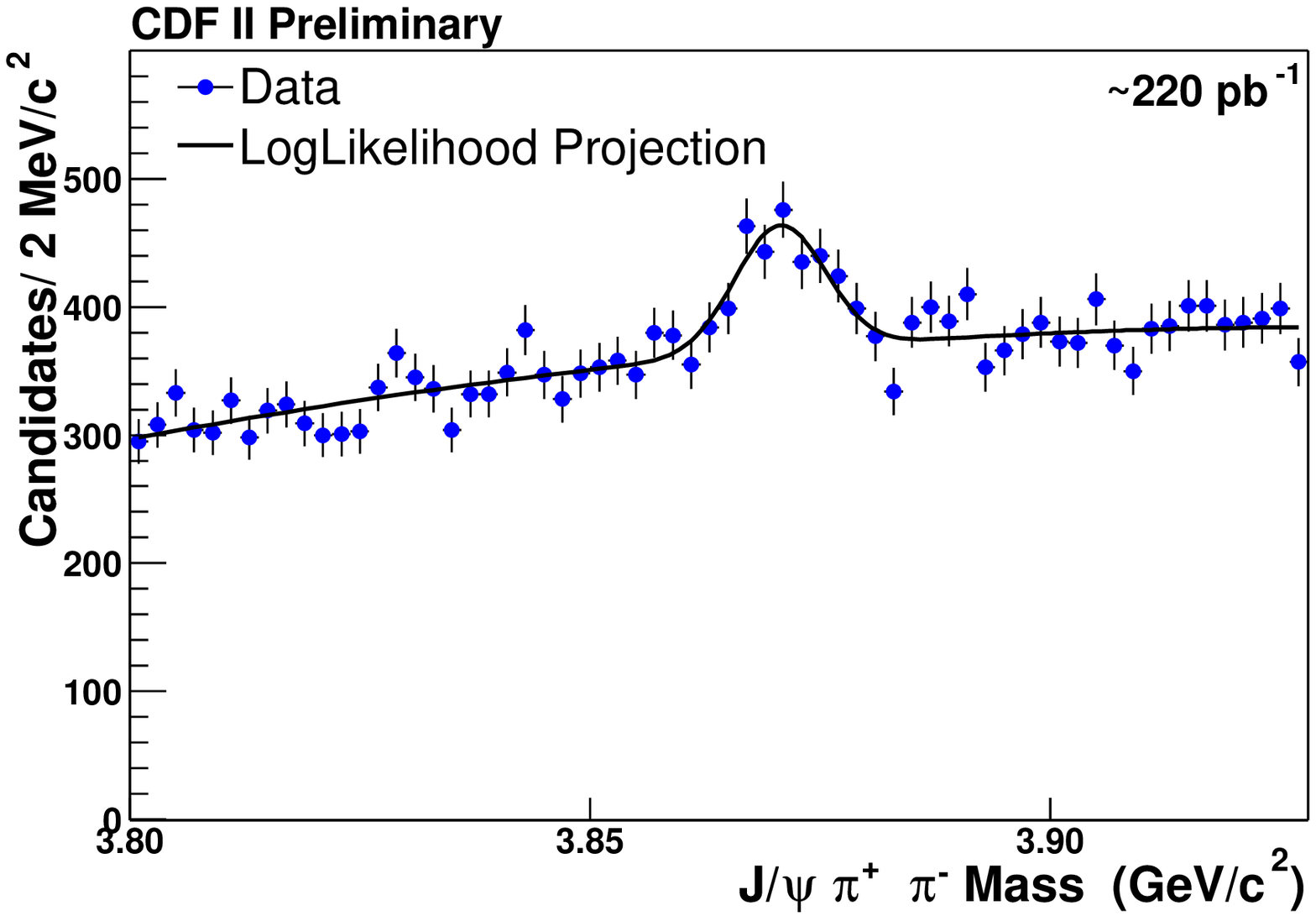,width=2.6in}\psfig{file=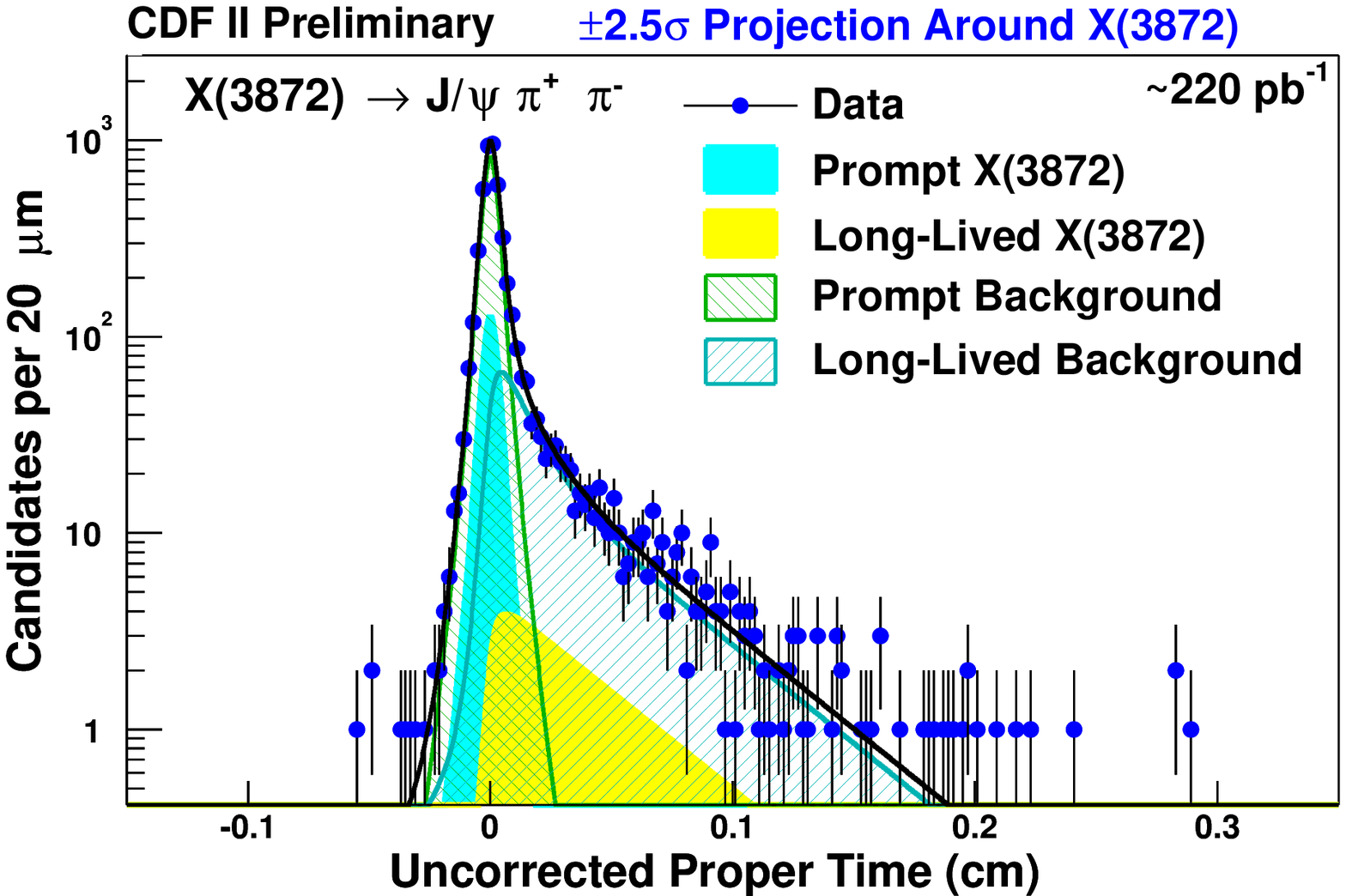,width=2.6in}
           }
\vspace*{8pt}
\caption{Projections of $X$-likelihood fit in 
         mass (left), and uncorrected 
         proper-time (right) as Fig.~\ref{Fig:2SCDF}.
\label{Fig:XCDF2}
}
\end{figure}

The fit for $\psi(2S)$ is shown in Fig.~\ref{Fig:2SCDF},
where $28.3\!\pm\! 1.0 \!\pm\! 0.7\%$ of  signal is displaced,
similar to Run~I results.\cite{Psi2SCDF}
For $X(3872)$,  with $M_{\pi\pi}\!>\!500\MeVcc$,
the fraction is $16.1\!\pm\! 4.9 \!\pm\! 2.0\%$ 
 (Fig.~\ref{Fig:XCDF2})---a bit 
more than $2\sigma$  from the  $\psi(2S)$.
These fractions agree  with those obtained by simple sideband subtraction.
They are, however, uncorrected for efficiency, 
and must be considered sample specific.\cite{Psi2SCDF}
The {\it absence} of a $b$-component is excluded at  $3\sigma$
based on  Monte Carlo ``experiments.''
%Data excludes the {\it absence} of a $b$-component at $3\sigma$
%from Monte Carlo ``experiments.''
%
Thus our $X$-sample is mainly prompt---presumably direct production---with 
a modest $b$-contribution.

It has been argued
that all conventional $c\bar{c}$ assignments
for the $X(3872)$ are problem\-atic.\cite{OlsenCharmAssign}
However, 
production of the $X$ appears, so far, quite similar to that of the 
$\psi(2S)$ in CDF.
If it is indeed a ``molecule,'' there seems to be no dramatic
penalty for producing such a fragile state in $\bar{p}p$ collisions.
Although, more incisive comparisons require specific theoretical models
for the production of exotic states.
A recent analysis of $X$-production as a $1^{++}$ state\cite{Braaten}
may benefit from our results.

Studies of this mysterious state are continuing in CDF.

% \section*{Acknowledgments}
% I would like to thank \ldots for inciting/provoking\ldots

% \begin{figure}[t]
% \centerline{\psfig{file=Blessed2S-TailProj.eps,width=2.9in}\psfig{file=BlessedX-TailProj.eps,width=2.8in}
%            }
% \vspace*{8pt}
% \caption{CDF  2S \& X(3872)
% \label{Fig:2SXCDFTail}
% }
% \end{figure}


\begin{thebibliography}{0}

\bibitem{BelleX}
  K. Abe  {\it et al.} (Belle), contribution to Lepton-Photon [hep-ex/0308029],
and reported in:
  T. Skwarnicki,  {\it 21st Int. Symp. On Lepton
    And Photon Interactions At High Energies},  August 2003, Fermilab,
  Int. J. Mod. Phys. {\bf A19},  1030 (2004) [hep-ph/0311243].


\bibitem{BelleXPRL}
  S.-L. Choi  {\it et al.} (Belle),   Phys. Rev. Lett. {\bf 91}, 262001 (2003).


% \bibitem{XCDFQuarkOnia} 
% G.Bauer  (CDF II),  {\it 2nd International Workshop on Heavy Quarkonium},
% 20-22 September 2003, Fermilab [web??]
%
 

\bibitem{XCDF}
  D. Acosta  {\it et al.} (CDF II),  Phys. Rev. Lett. {\bf 93}, 072001  (2004).


\bibitem{Yan}
T.M.~Yan, Phys. Rev. D {\bf 22}, 1652 (1980).

\bibitem{XLit}
An extensive list of literature on the $X(3872)$  can be found 
in the citations of Ref.~\refcite{Braaten}.

\bibitem{CDFII}
R. Blair  {\it et al.} (CDF II), FERMILAB-PUB-96-390-E (1996);
and citations in Ref.~\refcite{XCDF}.

% CDF - Run II Status and Prospects 
% M. Paulini, The CDF Collaboration, FERMILAB-CONF-03/010-E. 
% Published Proceedings Beauty 2002: 8th International Conference on B Physics at Hadron Machines,
% Santiago de Compostela, Spain, June 17-21, 2002.
% 

% \bibitem{XD0}
%   V.M. Abazov  {\it et al.} (D\O),
%   submitted to Phys. Rev. Lett.,
%   hep-ex/0405004.

\bibitem{Psi2SCDF}
  F. Abe  {\it et al.} (CDF),  Phys. Rev. Lett. {\bf 79}, 572  (1997).

\bibitem{OlsenCharmAssign}
S. Olsen (Belle), 
MESON 2004,
%8th Int. Workshop on Meson Production, Properties and Interactions, 
Cracow, Poland,  June 2004 
%MESON2004 Workshop  Krakow, June 2004 
[hep-ex/0407033].

\bibitem{Braaten}
E. Braaten [hep-ph/0408230].
%\\ 100


\end{thebibliography}
\end{document}